\documentclass[a4paper,11pt]{article}
\pdfoutput=1
\usepackage{jheppub}

\usepackage{amssymb,amsmath}
\usepackage[normalem]{ulem}
\usepackage[utf8x]{inputenc}
\usepackage{slashed}
\usepackage{graphicx}
\usepackage{here}
\usepackage{color}
\usepackage{csquotes} %fix backwards quotes
\usepackage{comment}
\usepackage{mathrsfs}
\usepackage{float}
\usepackage{ascmac}
\usepackage{multirow}
\usepackage{longtable}
\usepackage{bm}

\usepackage[italicdiff]{physics}
\usepackage[hang,small,bf]{caption}
\usepackage[subrefformat=parens]{subcaption}
\makeatletter
\newcommand*\rel@kern[1]{\kern#1\dimexpr\macc@kerna}
\newcommand*\widebar[1]{%
  \begingroup
  \def\mathaccent##1##2{%
    \rel@kern{0.8}%
    \overline{\rel@kern{-0.8}\macc@nucleus\rel@kern{0.2}}%
    \rel@kern{-0.2}%
  }%
  \macc@depth\@ne
  \let\math@bgroup\@empty \let\math@egroup\macc@set@skewchar
  \mathsurround\z@ \frozen@everymath{\mathgroup\macc@group\relax}%
  \macc@set@skewchar\relax
  \let\mathaccentV\macc@nested@a
  \macc@nested@a\relax111{#1}%
  \endgroup
}
\makeatother

\numberwithin{equation}{section}

\preprint{
\begin{minipage}{5cm}
\small
\flushright
EPHOU-24-013\\
KYUSHU-HET-295
\end{minipage}}

\title{Yukawa textures from non-invertible symmetries}

\author{Tatsuo Kobayashi$^{1}$,} 
\author{Hajime Otsuka$^{2}$, and} 
\author{Morimitsu Tanimoto$^{3}$} 
\affiliation{
$^1$Department of Physics, Hokkaido University, Sapporo 060-0810, Japan}
\affiliation{
$^2$Department of Physics, Kyushu University, 744 Motooka, Nishi-ku, Fukuoka 819-0395, Japan}
\affiliation{$^3$Department of Physics, Niigata University, Ikarashi 2-8050, Niigata 950-2181, Japan}
\emailAdd{kobayashi@particle.sci.hokudai.ac.jp}
\emailAdd{otsuka.hajime@phys.kyushu-u.ac.jp}
\emailAdd{morimitsutanimoto@yahoo.co.jp}

\abstract{
Phenomenological aspects of non-invertible symmetries, in particular the flavor structure of quarks and leptons, are studied.
We start with a $\mathbb{Z}_M$ discrete symmetry and gauge $\mathbb{Z}_2$ so as to obtain a non-invertible symmetry.
We study which Yukawa textures can be derived from the non-invertible symmetries.
Various textures can be realized and some of them cannot be realized by a conventional symmetry. 
For example, the nearest neighbor interaction texture
 as well as other interesting textures of quarks and leptons are obtained.
}

\makeatletter
\gdef\@fpheader{}
\makeatother

\begin{document}

\maketitle

\section{Introduction}

The origin of the flavor structure is one of the mysteries in particle physics.
Various approaches have been studied to understand the hierarchical fermion masses, 
large and small mixing angles, and CP-violating phases.

One type of approaches is to impose symmetries including Abelian and non-Abelian symmetries, and 
continuous and discrete ones. 
The $U(1)$ Froggatt-Nielsen mechanism is one of the famous ones to explain the hierarchy among quark 
and lepton masses, and their mixing angles \cite{Froggatt:1978nt}. 
Also non-Abelian discrete flavor symmetries have been studied very intensively. (See Refs. \cite{Altarelli:2010gt,Ishimori:2010au,Hernandez:2012ra,King:2013eh,King:2014nza,Petcov:2017ggy,Kobayashi:2022moq} for reviews.)
Recently, the modular flavor symmetries have been attracting much attention \cite{Feruglio:2017spp}. (See Refs. \cite{Kobayashi:2023zzc,Ding:2023htn} for reviews.)
In contrast to the conventional approach, the Yukawa couplings described by modular forms transform under the modular symmetry, and they are non-trivial representations of finite modular groups such as $S_3$, $A_4$, $S_4$, and $A_5$ \cite{Kobayashi:2018vbk,Feruglio:2017spp,Penedo:2018nmg,Novichkov:2018nkm}. 
Such a phenomenon naturally appears when the modular symmetry is regarded as a geometrical symmetry of compact space. 
Indeed, the modular flavor symmetries can be realized by extra-dimensional theories, e.g., heterotic orbifold models \cite{Ferrara:1989qb,Lerche:1989cs,Lauer:1989ax,Lauer:1990tm}, heterotic Calabi-Yau compactifications \cite{Ishiguro:2020nuf,Ishiguro:2021ccl,Ishiguro:2024xph}, and magnetized compactifications of type II string theory \cite{Kobayashi:2018rad,Kobayashi:2018bff,Ohki:2020bpo,Kikuchi:2020frp,Kikuchi:2020nxn,
Kikuchi:2021ogn,Almumin:2021fbk}.

Another type of approaches is to assume the texture of quark and lepton mass matrices, which was proposed by Weinberg  \cite{Weinberg:1977hb}
and Fritzsch \cite{Fritzsch:1977vd,Fritzsch:1979zq} at first. 
Some texture patterns can be realized by imposing certain symmetries, 
but others are difficult to be realized by symmetries in a simple way.

Recently, the concept of symmetries has been generalized.
In particular, various types of non-invertible symmetries were studied in many topics. (See for reviews about non-invertible symmetries in various dimensions, e.g., Refs.~\cite{Gomes:2023ahz,Schafer-Nameki:2023jdn,Bhardwaj:2023kri,Shao:2023gho}.) 
In the context of the Standard Model, the Abelian chiral symmetries lead to a non-invertible symmetry \cite{Choi:2022jqy,Cordova:2022ieu}. 
Furthermore, it was proposed in Ref. \cite{Cordova:2022fhg} that small Dirac neutrino Yukawa couplings are protected by a non-invertible symmetry, e.g., in a $U(1)_{L_\mu - L_\tau}$ gauge theory, and in addition non-invertible Peccei-Quinn symmetries were discussed as a solution to the strong CP problem \cite{Cordova:2024ypu}. 
Concerning the flavor symmetry of quarks and leptons, non-invertible symmetries was studied in Ref.~\cite{Kobayashi:2024yqq} within the framework of low-energy effective field theory derived from magnetized D-brane models of 
type IIB superstring theory on toroidal orbifold backgrounds. 
The $\mathbb{Z}_M$ symmetry among zero modes can appear in magnetized torus compactifications, where 
$M$ depends on the size of magnetic flux in the compact space. 
The orbifolding breaks the $\mathbb{Z}_M$ symmetry.
Any invertible symmetry does not remain for $M=$ odd, while the $\mathbb{Z}_2$ symmetry remains for $M=$ even. 
However, the allowed couplings are controlled by a certain selection rule even for $M=$ odd. 
That is the non-invertible symmetry. The coupling selection rule can be explained intuitively as follows. 
Charges $q$ and $M-q$ of the original $\mathbb{Z}_M$ symmetry are identified and one state has both charges. 
Hence, this is not an invertible symmetry.

Our purpose of this paper is to apply the non-invertible symmetry in Ref.~\cite{Kobayashi:2024yqq} 
to the bottom-up approach of flavor model building and then to study which pattern of mass matrices can be realized. 
As a result, we show that we can derive 
{\it the nearest neighbor interaction} (NNI) texture \cite{Branco:1988iq,Branco:1994jx,Branco:1999nb} and other textures, which cannot be realized by the conventional symmetry approach.

This paper is organized as follows.
In section \ref{sec:non-invertible}, we explain the non-invertible symmetry, which we use in this paper.
In section \ref{sec:texture}, we show which texture can be realized by our non-invertible symmetry.
In section \ref{sec:phenomenology}, we discuss phenomenological implications of our results.
Section \ref{sec:con} is devoted to the conclusion.

%%%%%%%%%%%%%%%%%%%%%%%%%%%%%%%%%%%%%%%%%%%%%%%%%%%%%%%%%%%%%%%%%
\section{Non-invertible symmetries}
\label{sec:non-invertible}
%%%%%%%%%%%%%%%%%%%%%%%%%%%%%%%%%%%%%%%%%%%%%%%%%%%%%%%%%%%%%%%%%

Let us start with the $\mathbb{Z}_M$ symmetry whose generators are represented by $g$. \footnote{This $\mathbb{Z}_M$ is a geometric symmetry derived from compact spaces such as string theory. 
In particular, the $\mathbb{Z}_M$ symmetry is originated from a discrete translation on the torus \cite{Abe:2009vi,Berasaluce-Gonzalez:2012abm,Marchesano:2013ega}.}
Its conjugacy classes are given by 
\begin{align}
g^k\qquad (k=0,1,\cdots,M-1).
\end{align}
These conjugacy classes correspond to representations with $M$ kinds of $\mathbb{Z}_M$ charges, which distinguish the states.
Obviously, the product satisfies
\begin{align}
g^{k_1}g^{k_2}=g^{k_1+k_2}.
\end{align}
If the right side is of the same conjugacy class as $g^0$,
it allows a two-point coupling between two states with certain $\mathbb{Z}_M$ charges. 
The same is true for $n$-point couplings, which can be allowed when the following condition is satisfied
\begin{align}
g^{k_1}g^{k_2}\cdots g^{k_n}=g^0.
\end{align}

Next, we consider the following automorphism to construct a non-invertible symmetry:
\begin{align}
ege^{-1} = g, \qquad rgr^{-1}= g^{-1}.
\end{align}
Using this, we define the following class
\begin{align}
[g^k]=\{ hgh^{-1} ~|~ h=e,r \}.
\end{align}
The idea of this class is based on a non-invertible symmetry, 
where the $\mathbb{Z}_2$ symmetry associated with $r$ is gauged.  \footnote{It corresponds to consider $T^2/\mathbb{Z}_2$ for an extra-dimensional compact space in string theory, where $\mathbb{Z}_2$ symmetry is gauged}.

For $M=2p$ and $M=2p+1$, there exist $p+1$ kinds of classes, where $k$ runs from $0$ to $p$.\footnote{The number of classes is exactly the same as the number of zero modes on the $T^2/\mathbb{Z}_2$ compactification with the magnetic flux $M$ \cite{Abe:2008fi}.} 
These classes distinguish the representations of the non-invertible symmetry, and the states have these representations.

The product of the classes is obtained as\footnote{Note that $[g^{k+k'}]$ ($[g^{M-k+k'}]$) belongs to the same class as $[g^{2M-k-k'}]$ ($[g^{M+k-k'}]$).} 
\begin{align}
[g^k][g^{k'}]=[g^{k+k'}]+[g^{M-k+k'}].
\end{align}
If the class of $[g^0]$ appears in the right-hand side, it 
allows two-point couplings between a certain state. 
The same is true for $n$-point couplings.

Intuitively, the selection rules can be summarized as follows. 
A state has a representation corresponding to a class $[g^k]$.
In terms of the original $\mathbb{Z}_M$ charge, 
the state has both the $k$ charge and $M-k$ charge simultaneously.
The other class $[g^{k'}]$ is similarly 
having $k'$ charge and $M-k'$ charge at the same time. 
The two-point coupling (mass term) of this state is 
allowed when
\begin{align}
\pm k \pm k'=0 \quad ({\rm mod}~M)
\end{align}
is satisfied. 
However, this coupling is allowed only for the same class $[g^k]=[g^{k'}]$. 
It indicates that the mass term (without insertion of the Higgs field vacuum expectation value) is always diagonal. 
Also, this argument can be applied to the kinetic terms, 
which are also non-zero in the same class $[g^k]=[g^{k'}]$, 
and there is no mixing between different classes.

The same argument can be discussed for a 3-point coupling. 
The conditions under which the 3-point coupling among classes $[g^k]$, $[g^{k'}]$ and $[g^{k''}]$ is allowed,  are given by
\begin{align}
\pm k \pm k' \pm k''=0 \quad ({\rm mod}~M).
\end{align}
In the case of the Yukawa coupling, the Yukawa matrix is not necessarily diagonal, depending on the class of the Higgs field. 
In the following section, we show the case where the Yukawa texture is indeed non-trivial.

It is straightforward to extend the previous discussion to a general $n$-point coupling. 
The conditions to have a coupling among the class $g^{k_i}$ ($i=1,2,\cdots, n$) are given by
\begin{align}
\sum_i \pm k_i = 0 \quad ({\rm mod}~M).
\end{align}

%%%%%%%%%%%%%%%%%%%%%%%%%%%%%%%%%%%%%%%%%%%%%%%%%%%%%%%%%%%%%%%%%
\section{Texture from non-invertible symmetry}
\label{sec:texture}
%%%%%%%%%%%%%%%%%%%%%%%%%%%%%%%%%%%%%%%%%%%%%%%%%%%%%%%%%%%%%%%%%

In this section, we present explicit examples. 
As studied in the previous section, we start with the $\mathbb{Z}_M$ symmetry, and further gauge $\mathbb{Z}_2$. 
Then, the fields $\phi$ including fermions and Higgs fields have representations of the non-invertible symmetry corresponding to the class $[g^k]$ as studied in the previous section, i.e., $\phi_{[g^k]}$.
Intuitively, the fields have 
the $k$ and $M-k$ at the same time, and the coupling selection rule is controlled by 
$\mathbb{Z}_M$.
Here, we study Yukawa textures derived from this non-invertible symmetry.

For $M=2$, after gauging $\mathbb{Z}_2$, 
the ``invertible'' $\mathbb{Z}_2$ flavor symmetry remains.
Hence, we deal with $M=3,4,5,6,7$ cases.

%%%%%%%%%%%%%%%%%%%%%%%%%%%%%%%%%%%%%%%%%%%%%%%%%%%%%%%%%%%%%%%%%
\subsection{$M=3$}
%%%%%%%%%%%%%%%%%%%%%%%%%%%%%%%%%%%%%%%%%%%%%%%%%%%%%%%%%%%%%%%%%

We start with the $M=3$ case. 
In this case, there are two classes to distinguish states: $[g^0]$ and $[g^1]$, each which corresponds to three generations of left- and right-handed fermions.  

When the representation of the Higgs field corresponds to $[g^0]$, the Yukawa matrix is described by
\begin{align}
Y_{[g^0]}=
\begin{pmatrix}
a & 0 \\
0 & b 
\label{M30}
\end{pmatrix}
.
\end{align}
Throughout this paper, we denote by $a,b,c,...,g$ complex numbers unless specified otherwise. 
The ordering of the fermions is $[g^0]$, $[g^1]$ for both rows and columns. 
As a result, this case leads to a diagonal matrix.
The result holds for a conventional $\mathbb{Z}_3$ symmetry as well as a $\mathbb{Z}_2$ symmetry. 
In the following, we will discuss the case where the Higgs field has the different representation, $[g^1]$, but the order of the generations is still the same for both row and column, i.e., $[g^0]$, $[g^1]$. 
In this case, we find
\begin{align}
Y_{[g^1]}=
\begin{pmatrix}
0 & a \\
b & c
\label{M31}
\end{pmatrix}
.
\end{align}
Obviously, we can not derive this pattern 
by assuming $U(1)$ symmetry or a conventional $\mathbb{Z}_M$ for any $M$.
Suppose that two left-handed (right-handed) fermions have $U(1)$ $Q_1$ and $Q_2$ ($q_1$ and $q_2$).
Non-vanishing (1,2), (2,1), and (2,2) entries require
\begin{align}
    Q_1+q_2=Q_2+q_1=Q_2+q_2,
\end{align}
where this sum may correspond to the $U(1)$ charge of the Higgs fields.
This equation leads to $Q_1=Q_2$ and $q_1=q_2$, but 
that is not consistent with the forbidden (1,1) entry.
Thus, there is no solution of $U(1)$ charges leading to the above pattern.
Also, this pattern is not consistent with 
a conventional $\mathbb{Z}_M$ symmetry for any $M$.

For $M=3$, we have just two class: $[g^0]$ and $[g^1]$.
Two generations among three generations of fermions are degenerate in the class $[g^k]$.
When $M=3$, Yukawa matrix of three generations can be obtained by 
\begin{align}
    Y=
    \begin{pmatrix}
        a & 0 & 0 \\
         0 & b & c \\
        0 & d & e   
    \end{pmatrix}
\end{align}
including possible permutations of rows and columns, or 
\begin{align}
    Y=
    \begin{pmatrix}
        0 & a & b \\
        c & d & e \\
        f & g & h    
    \end{pmatrix}
\end{align}
including possible permutations of rows and columns.
The former has 4 textures zeros, but this pattern 
can be derived by a conventional Abelian symmetry.
The latter has one texture zero, and this pattern cannot be derived by a conventional symmetry.

%%%%%%%%%%%%%%%%%%%%%%%%%%%%%%%%%%%%%%%%%%%%%%%%%%%%%%%%%%%%%%%%%
\subsection{$M=4$}
%%%%%%%%%%%%%%%%%%%%%%%%%%%%%%%%%%%%%%%%%%%%%%%%%%%%%%%%%%%%%%%%%

Let us examine the $M=4$ case. 
In this case, there are three classes to distinguish the states: $[g^0]$, $[g^1]$, and $[g^2]$, which correspond to three generations of left- and right-handed fermions.  

When the representation of the Higgs field corresponds to $[g^0]$, the Yukawa matrix is described by
\begin{align}
Y_{[g^0]}=
\begin{pmatrix}
a & 0 & 0\\
0 & b & 0 \\
0 & 0 & c
\end{pmatrix}
.
\end{align}
The ordering of the generations is $[g^0]$, $[g^1]$, and $[g^2]$ for both rows and columns. 
The selection rule of the non-invertible symmetry leads to a diagonal matrix.
The result holds for a conventional $\mathbb{Z}_4$. 
In the following, we will discuss the case where the Higgs field has a different representation such as $[g^1]$ and $[g^2]$, but the order of the generations is still the same for both rows and columns, i.e., $[g^0]$, $[g^1]$, and $[g^2]$.
Yukawa matrices can be written by
\begin{align}
Y_{[g^1]}=
\begin{pmatrix}
0 & a & 0\\
b & 0 & c \\
0 & d & 0
\end{pmatrix}
,\qquad
Y_{[g^2]}=
\begin{pmatrix}
0 & 0 & a\\
0 & b & 0 \\
c & 0 & 0
\end{pmatrix},
\end{align}
when the Higgs field corresponds to 
$[g^1]$ and $[g^2]$, respectively.

The pattern $Y_{[g^2]}$ is equivalent to 
$Y_{[g^0]}$ by exchanging rows and columns. 
For $Y_{[g^1]}$, the restriction is more severe than $\mathbb{Z}_2$ symmetry, and some elements are zero.
For example, the (1,3) and (3,1) components of $Y_{[g^1]}$ are allowed in $\mathbb{Z}_2$ symmetry, but not allowed in this non-invertible symmetry.

%%%%%%%%%%%%%%%%%%%%%%%%%%%%%%%%%%%%%%%%%%%%%%%%%%%%%%%%%%%%%%%%%
\subsection{$M=5$}
%%%%%%%%%%%%%%%%%%%%%%%%%%%%%%%%%%%%%%%%%%%%%%%%%%%%%%%%%%%%%%%%%

In this case, there are three different representations: $[g^0]$, $[g^1]$, and $[g^2]$, which correspond to three generations of left- and right-handed fermions. 
The notation is the same as above.

When the representation of the Higgs field corresponds to $[g^0]$, the Yukawa matrix is given by
\begin{align}
Y_{[g^0]}=
\begin{pmatrix}
a & 0 & 0\\
0 & b & 0 \\
0 & 0& c
\label{M5-0}
\end{pmatrix}
.
\end{align}
The ordering of the generations is $[g^0]$, $[g^1]$, $[g^2]$ for both rows and columns. 
As a result, this case leads to a diagonal matrix.
The result holds for a conventional $\mathbb{Z}_5$ or $\mathbb{Z}_3$. 
Although we will discuss the case where the Higgs field has other representations,
the ordering of the generations is still $[g^0]$, $[g^1]$, $[g^2]$ for both rows and columns.

Next, we deal with the case where the representation of the Higgs field is given by $[g^1]$, which leads to the following configuration of the Yukawa matrix:
\begin{align}
Y_{[g^1]}=
\begin{pmatrix}
0 & a & 0\\
b & 0 & c \\
0 & d & e
\label{M5-1}
\end{pmatrix}
.
\end{align}
Remarkably, this reproduces the NNI form \cite{Branco:1988iq,Branco:1994jx,Branco:1999nb}.  
The right-bottom $(2 \times 2)$ submatrix is the same as $Y_{[g^1]}$ for $M=3$, 
which cannot be simply realized by a 
conventional symmetry.
Similarly, one can not derive the NNI type 
simply by a conventional symmetry.
Then, it is necessary to extend models by introducing new particles such as multi-Higgs. 
(See, for example, Ref.~\cite{Kikuchi:2022svo}.)
Hence, the non-invertible symmetry allows us to derive mass matrices that cannot be derived by the conventional symmetry.

We move to the case where the representation of the Higgs field is $[g^2]$. The Yukawa matrix is described by
\begin{align}
Y_{[g^2]}=
\begin{pmatrix}
0 & 0 & a\\
0 & b & c \\
d & e& 0
\label{M5-2}
\end{pmatrix}
.
\end{align}
This pattern is also difficult to derive from the conventional symmetry argument. 
Note that there is a degree of freedom to change the order of the left-handed and right-handed fermions.

%%%%%%%%%%%%%%%%%%%%%%%%%%%%%%%%%%%%%%%%%%%%%%%%%%%%%%%%%%%%%%%%%
\subsection{$M=6$}
%%%%%%%%%%%%%%%%%%%%%%%%%%%%%%%%%%%%%%%%%%%%%%%%%%%%%%%%%%%%%%%%%

In this case, there are four types of representations, i.e., $[g^0]$, $[g^1]$, $[g^2]$, $[g^3]$. 
For the representation of Higgs field fixed as 
$[g^k]$, four left-handed and right-handed states can have the following patterns of couplings:
\begin{align}
Y_{[g^0]}=
\begin{pmatrix}
a & 0 & 0 & 0 \\
0 & b & 0 & 0 \\
0 & 0 & c & 0 \\
0 & 0 & 0 & d
\end{pmatrix}
,\quad
Y_{[g^1]}=
\begin{pmatrix}
0 & a & 0 & 0\\
b & 0 & c & 0\\
0 & d & 0 & e \\
0 & 0 & f & 0
\end{pmatrix}
,\quad
Y_{[g^2]}=
\begin{pmatrix}
0 & 0 & a & 0 \\
0 & b & 0 & c \\
d & 0 & e & 0 \\
0 & f & 0 & 0
\end{pmatrix}
,\quad
Y_{[g^3]}=
\begin{pmatrix}
0 & 0 & 0 & a \\
0 & 0 & b & 0 \\
0 & c & 0 & 0 \\
e & 0 & 0 & 0
\end{pmatrix}
.
\end{align}
The ordering of both rows and columns is 
$[g^0]$, $[g^1]$, $[g^2]$, $[g^3]$. 
The patterns, $Y_{[g^0]}$ and $Y_{[g^3]}$ are 
equivalent by permutations of rows and columns.
Also, $Y_{[g^1]}$ and $Y_{[g^2]}$ are equivalent by permutations.

In three-generation models, we pick up 
three generations from four rows and columns 
in the above matrices.
From $Y_{[g^0]}$ and $Y_{[g^3]}$, 
we can obtain $(3\times 3)$ diagonal matrix including its permutations and 
the matrices with $\det Y=0$.
From $Y_{[g^1]}$ and $Y_{[g^2]}$, 
we can obtain the following $(3 \times 3)$ 
Yukawa matrices:
\begin{align}
    Y=
    \begin{pmatrix}
        0 & a & 0 \\
        b & 0 & c \\
        0 & d & 0
    \end{pmatrix}, \quad
    Y= 
    \begin{pmatrix}
    a & 0 & b \\
    0 & c & 0 \\
    0 & 0 & d
    \end{pmatrix}
    ,
\end{align}
and their possible permutations of rows and columns 
as well as $(3 \times 3)$ Yukawa matrices 
with $\det Y=0$.
The former corresponds to $Y_{[g^1]}$ for $M=4$.

%%%%%%%%%%%%%%%%%%%%%%%%%%%%%%%%%%%%%%%%%%%%%%%%%%%%%%%%%%%%%%%%%
\subsection{$M=7$}
%%%%%%%%%%%%%%%%%%%%%%%%%%%%%%%%%%%%%%%%%%%%%%%%%%%%%%%%%%%%%%%%%

Here, the Yukawa matrices for $M=7$ are shown.
In this case, there are four types of representations, i.e., $[g^0]$, $[g^1]$, $[g^2]$, $[g^3]$. 
We assign three of these combinations to the left-handed and right-handed fermions with 3 generations. 
We present four kinds of the Yukawa matrix in the ordering of $[g^0]$, $[g^1]$,$[g^2]$, $[g^3]$ as follows:
\begin{align}
Y_{[g^0]}=
\begin{pmatrix}
a & 0 & 0 & 0 \\
0 & b & 0 & 0 \\
0 & 0 & c & 0 \\
0 & 0 & 0 & d
\end{pmatrix}
,\quad
Y_{[g^1]}=
\begin{pmatrix}
0 & a & 0 & 0\\\
b & 0 & c & 0\\
0 & d & 0 & e \\
0 & 0 & f & g
\end{pmatrix}
,\quad
Y_{[g^2]}=
\begin{pmatrix}
0 & 0 & a & 0 \\
0 & b & 0 & c\\
d & 0 & 0 & e \\
0 & f & g & 0
\end{pmatrix}
,\quad
Y_{[g^3]}=
\begin{pmatrix}
0 & 0 & 0 & a \\
0 & 0 & b & c \\
0 & d & e & 0 \\
f & g & 0 & 0
\end{pmatrix}
.
\end{align}
Notation is the same as the above, 
and we impose that the representations of Higgs fields are $[g^0]$, $[g^1]$, $[g^2]$ and $[g^3]$, respectively. 
$Y_{[g^1]}$ and $Y_{[g^3]}$ are equivalent by 
permutations of rows and columns.

In the three-generation models, we pick up three generations from four rows and columns in the above matrices.
From $Y_{[g^0]}$, we can obtain $(3 \times 3) $ diagonal matrix including its permutations of rows and columns and the matrices with $\det Y=0$.
From $Y_{[g^1]}$ and $Y_{[g^3]}$, 
we can obtain the following $(3 \times 3)$ Yukawa matrices:
\begin{align}
   & Y= \begin{pmatrix}
        0 & a & 0 \\
        b & 0 & c \\
        0 & d & e
    \end{pmatrix}, \quad
    Y=\begin{pmatrix}
        a & b & 0 \\
        0 & 0 & c \\
        0 & d & e
    \end{pmatrix}, \quad
    Y=\begin{pmatrix}
        a & 0 & 0 \\
        0 & b & c \\
        0 & d & e
    \end{pmatrix}, \quad
    Y=\begin{pmatrix}
        a & 0 & 0 \\
        b & 0 & c \\
        0 & d & e
    \end{pmatrix},  \notag \\
   & Y=\begin{pmatrix}
        a & 0 & 0 \\
        0 & b & 0 \\
        0 & c & d
    \end{pmatrix}, \quad
    Y=\begin{pmatrix}
        0 & a & 0 \\
        b & 0 & 0 \\
        0 & 0 & c
    \end{pmatrix}, \quad
    Y=\begin{pmatrix}
        0 & a & 0 \\
        b & 0 & c \\
        0 & d & 0
    \end{pmatrix},
\end{align}
and their possible permutations of rows and columns as well as $(3\times 3)$ Yukawa matrices with $\det Y=0$.
From $Y_{[g^2]}$, 
we can obtain the following $(3 \times 3)$ Yukawa matrices:
\begin{align}
   & Y= \begin{pmatrix}
        a & 0 & b \\
        0 & 0 & c \\
        d & e & 0
    \end{pmatrix}, \quad
    Y=\begin{pmatrix}
        0 & 0 & a \\
        b & 0 & c \\
        0 & d & 0
    \end{pmatrix},
\end{align}
and their possible permutations of rows and columns as well as $(3\times 3)$ Yukawa matrices with $\det Y=0$.

Similarly, we can study the cases with larger $M$.

%%%%%%%%%%%%%%%%%%%%%%%%%%%%%%%%%%%%%%%%%%%%%%%%%%%%%%%%%%%%%%%%%
\section{Phenomenological implications}
\label{sec:phenomenology}
%%%%%%%%%%%%%%%%%%%%%%%%%%%%%%%%%%%%%%%%%%%%%%%%%%%%%%%%%%%%%%%%%
As discussed in the previous section,
the non-invertible symmetry is favorable to the texture zeros of
the Yukawa matrices of quarks and leptons.
The texture zeros approach has a long history.
In the framework of two families of quarks, Weinberg considered a mass matrix for the down-type quark 
sector with zero  (1,1) entry in the basis in which the 
up-type quark mass matrix is diagonal \cite{Weinberg:1977hb}.  
Then the Cabibbo angle is successfully predicted to be $\sqrt{m_d/m_s}$.
which is the so-called Gatto, Sartori, Tonin relation \cite{Gatto:1968ss}. 
This case is realized easily as seen in
$Y_{[g^0]}$ and $Y_{[g^1]}$  of Eqs.~\eqref{M30} and \eqref{M31}
 in the case of $M=3$.
%%%%%%%%%%%%%%%%%%%%%%%%%%%%%%

Fritzsch extended the above approach to the three
family case \cite{Fritzsch:1977vd,Fritzsch:1979zq}. 
Ramond, Roberts and Ross presented 
a systematic analysis with  four or five zeros
for symmetric or hermitian quark mass matrices \cite{Ramond:1993kv}.
Their textures are not viable today since they cannot describe 
the current rather precise data on the CKM quark mixing 
matrix. However, the texture zero approach to the  mass matrices 
of quarks and leptons
is still promising \cite{Fritzsch:2002ga,Xing:2015sva,Frampton:2002yf,
Kageyama:2002zw}.
Also one can obtain 
some  sets of zeros of the quark mass matrices
by making a suitable weak basis transformation. 
This issue is 
well known as NNI  basis \cite{Branco:1988iq,Branco:1994jx,Branco:1999nb}.

In what follows, we study quark and lepton Yukawa matrices as well as mass matrices within the framework of supersymmetric models. That is because we would like to introduce the up-sector and down-sector Higgs fields, $H_U$ and $H_D$, with different representations $[g^k]$ in order to derive different Yukawa matrices between the down-type and up-type quarks, and neutrinos and charge leptons.\footnote{In this case, we need an additional singlet field to generate the $\mu$-term as in the Next-To-Minimal Supersymmetric Standard Model.}
Type II non-supersymmetric two doublet Higgs models would lead to the same patterns. 

%%%%%%%%%%%%%%%%%%%%%%%%
\subsection{Quark sector}
%%%%%%%%%%%%%%%%%%%%%%%%
In the NNI basis, the quark Yukawa matrices are given as 
\begin{align}
&  Y_D =
\begin{pmatrix}
0  &  a_D &0  \\
a'_D & 0 & b_D \\
0& b'_D& c_D
\end{pmatrix}_{LR}\,,
\qquad
Y_U =
\begin{pmatrix}
0  &  a_U &0  \\
a'_U & 0 & b_U \\
0& b'_D& c_U
\end{pmatrix}_{LR}\,,
\label{NNI-quark}
\end{align}
where the  coefficient of each element is complex in general
\footnote{Among ten phases, eight phases  are removed  
	by the redefinition of the quark fields. }. 
This texture of Yukawa matrices is found in  $Y_{[g^1]}$
of Eq.~\eqref{M5-1}. 
That is the case of  $M=5$ with Higgs representation $[g^1]$.
Since this basis is also obtained by choosing a suitable weak basis transformation from general $3\times 3$ Yukawa matrices 
of down-type and up-type quarks \cite{Branco:1988iq},
the  Yukawa matrices of Eq.~\eqref{NNI-quark} is completely consistent with observed CKM matrices and quark masses.
Thus, the non-invertible symmetry is compatible with the NNI basis.
On the other hand, the complicated set-up is required to obtain
the NNI basis in the conventional discrete flavor symmetry \cite{Kikuchi:2022svo}.

%%%%%%%%%%%%%%%%%%%%%%
%%%%%%%%%%%%%%%%%%%%%%%%%%%%%%
On the other hand,
a systematic study of texture zeros has been presented for the down-type  quark mass matrix in the basis of diagonal up-type quark mass matrix 
in Ref.~\cite{Tanimoto:2016rqy} from  
the standpoint of ``Occam's Razor approach'' \cite{Harigaya:2012bw},
in which a minimum number of parameters is  allowed.
The down-type quark mass matrix was arranged to have the minimum 
number of parameters by 
setting three of its elements to zero, while at the 
same time requiring that it
describes successfully the CKM mixing and CP violation without 
assuming it to be symmetric or hermitian.

We easily obtain a set of quark mass matrices: 
\begin{align}
Y_D=
\begin{pmatrix}
0 & 0 & a_D  \\
a'_D & c_D & b_D \\
c'_D & d_D & 0
\end{pmatrix}_{LR}\,,\qquad 
Y_U =
\begin{pmatrix}
a_U &0 &0  \\
0 & b_U  & 0\\
0& 0& c_U
\end{pmatrix}_{LR},
\label{texture-Occam1}
\end{align}
where $M=5$ with  the representation of the Higgs $H_D$, three generations of left-handed quarks, three generations of right-handed quarks for the down-type quarks being respectively assigned as
$[g^2]$, $\{[g^0], [g^1], [g^2]\}$, $\{[g^1], [g^1], [g^2]\}$, while the representation of the Higgs $H_U$,  and three generations of right-handed quarks for the up-type quarks are respectively assigned as
$[g^0]$, $\{[g^0], [g^1], [g^2]\}$. 
The texture of Eq.\,\eqref{texture-Occam1}
is equivalent to $M_d^{(7)}$ in the Appendix A of  Ref.~\cite{Tanimoto:2016rqy}.
However, this texture is excluded by the observed CKM mixing 
element $V^{\rm CLM}_{ub}$ in the strict sense
although it predicts  $V^{\rm CKM}_{ub}={\cal O}(\lambda^3)$
consistent with the order of the observation
where $\lambda$ is the Cabibbo angle.

The other one is:
\begin{align}
&  Y_D=
\begin{pmatrix}
 a_D &  a'_D &0\\
0 & b_D  & c_D \\
0& c'_D& d_D
\end{pmatrix}_{LR}\,,
\qquad
Y_U =
\begin{pmatrix}
a_U &0 &0  \\
0 & b_U  & c_U\\
0& c'_U& d_U
\end{pmatrix}_{LR},
\label{}
\end{align}
where $Y_D$ corresponds to $M_d^{(2)}$
 in  Ref.~\cite{Tanimoto:2016rqy}. 
 The texture of $Y_D$ is obtained in
 the case with $M=5$ by setting the representation of the Higgs
 $H_D$, three generations of left-handed quarks, three generations of right-handed quarks $[g^2]$, $\{[g^1], [g^2], [g^2]\}$, $\{[g^2], [g^1], [g^0]\}$, respectively. 
  On the other hand, $Y_U$ is obtained
  by taking the representation of the Higgs
 $H_U$ being $[g^0]$ with the same assignments of up-type quarks as the down-type quarks then,
  $Y_U$ is  not diagonal because of
  the degeneracy of the representation of the second and third generations.
  Since $|d_U|\gg |c_U|$ due to quark mass hierarchy, this set works well 
  in the experimental data of the CKM matrix.

%%%%%%%%%%%%%%%%%%%%%%%%%%%%%%%
%%%%%%%%%%%%%%%%%%%%%%%%%%%%%%%
%%%%%%%%%%%%%%%%%%%%%%%%%%%%%%%

\subsection{Lepton sector}
We discuss the texture zeros in the lepton sector.
 The neutrino mass matrix was investigated in the framework of the seesaw mechanism based on the Occam's Razor approach
 \cite{Kaneta:2016gbq}. Imposing four zeros in the Dirac neutrino Yukawa matrix gives the minimum number of parameters needed for the observed neutrino masses
and lepton mixing angles, while the charged lepton Yukawa matrix and the right-handed Majorana neutrino mass matrix are diagonal ones. The low-energy neutrino mass matrix has only seven physical parameters. Among them, the  CP phases are two Majorana phases which appear in the Majorana mass matrix.
 The matrices are given as:
%%%%%%%%%%%%%%
\begin{align}
&  Y_E =
\begin{pmatrix}
 a_E &0 & 0 \\ 0 & b_E& 0\\ 0& 0& c_E
\end{pmatrix}_{LR}\hskip -2mm,
\quad
&  Y_{\nu D} =
\begin{pmatrix}
0  &  a_\nu &0  \\
a'_\nu&0& c_\nu \\
0& c'_\nu& d_\nu
\end{pmatrix}_{LR}\hskip -2mm,
\quad
M_R =
\begin{pmatrix}
M_{1}\, e^{i \alpha} &0 &0  \\
0 & M_{2}\, e^{i \beta}  & 0\\
0& 0& M_{3}
\end{pmatrix},
\label{texture-lepton}
\end{align}
where all parameters are real.
%%%%%%%%%%%%%%%%%%%%%%%%%%%%%%%%%

Those textures are obtained in the non-invertible symmetry.
The Dirac neutrino texture is  $Y_{[g^1]}$
in  Eq.~\eqref{M5-1}, where $M=5$ with Higgs $H_U$ representation $[g^1]$.
On the other hand, the diagonal charged leptons 
are given by putting  $M=5$ with Higgs $H_D$ representation $[g^0]$.
The right-handed Majorana mass matrix is guaranteed to be diagonal.

%%%%%%%%%%%%%%%%%%%%%%%%%%%%%%%%%
The texture is completely consistent with the observed
neutrino mass squared differences and 
three mixing angles with the normal mass hierarchy.
The CP-violating Dirac phase is also consistent with recent data
of NuFIT5.3 (2024)  \cite{Esteban:2020cvm}.

%%%%%%%%%%%%%%%%%%%%%%%%%%%%%%%%%%%%%%%%%%%%%%%%%%%%%%%%%%%%%%%%%
\section{Conclusions}
\label{sec:con}
%%%%%%%%%%%%%%%%%%%%%%%%%%%%%%%%%%%%%%%%%%%%%%%%%%%%%%%%%%%%%%%%%

We have studied phenomenological aspects of non-invertible symmetries, in particular the flavor structure
of quarks and leptons.
We started with the $\mathbb{Z}_M$ discrete symmetry and gauged the $\mathbb{Z}_2$ symmetry so as to obtain the non-invertible symmetry.
We have studied which patterns of Yukawa matrices can be derived from the non-invertible symmetry.
As a result, we can realize various Yukawa textures.
Some of them cannot be derived by a conventional invertible symmetry in a simple way.
For example, we can realize the NNI texture and other interesting textures including the textures in ``Occam's Razor approach'' \cite{Harigaya:2012bw} can be obtained. 
We have discussed the phenomenological implications of our results on quark and lepton mass matrices. 
Thus, the non-invertible symmetry is quite important in flavor physics.

It is important to extend our analysis to other flavor aspects by the non-invertible symmetries, e.g. higher dimensional operators in flavor physics.
Also it is important to study phenomenological aspects of non-invertible symmetries other than the symmetry, which we have studied here.
Our non-invertible symmetry can be originated from string compactifications.
It is interesting to study which compactifications lead to the NNI texture and other interesting textures. 
In this case, we may consider other $\mathbb{Z}_N$ orbifoldings which will lead to the different Yukawa textures. 
Relevant studies will appear elsewhere.

\acknowledgments

This work was supported in part JSPS KAKENHI Grant Numbers JP23H04512 (H.O) and JP23K03375 (T.K.).

\appendix

\bibliography{references}{}
\bibliographystyle{JHEP}

\end{document}